\documentclass{ws-procs975x65}            % default, citations in superscript
\pdfoutput=1                                              % use PDFlatex to compile
\begin{document}
\title{From the Fourth Color to Spin-charge Separation --- \\ Neutrinos and Spinons}

\author{CHI XIONG$^*$}

\address{Institute of Advanced Studies, Nanyang Technological University,\\
60 Nanyang View, Singapore 639673, Singapore\\
$^*$E-mail: xiongchi@ntu.edu.sg}

\begin{abstract}
We introduce the spin-charge separation mechanism to the quark-lepton unification models which consider the lepton number as the fourth color. In certain finite-density systems, quarks and leptons are decomposed into spinons and chargons, which carry the spin and charge degrees of freedom respectively. Neutrinos can be related to the spinons with respect to the electric-charge and spin separation in the early universe or other circumstances. Some effective, probably universal couplings between the spinon sector and the chargon sector are derived and a phenomenological description for the chargon condensate is proposed. It is then demonstrated that the spinon current can induce vorticity in the chargon condensate, and spinon zero modes are trapped in the vortices, forming spinon-vortex bound states. In cosmology this configuration may lead to the emission of extremely high energy neutrinos when vortices split and reconnect.    
\end{abstract}


\bodymatter

%%%%%%%%%%%%%%%%%%%%%%%%%%%%%%%%%%%%%%%
\section{The fourth color and the haplon model}
%%%%%%%%%%%%%%%%%%%%%%%%%%%%%%%%%%%%%%%

Neutrinos are one of the most important probes to the quark-lepton unification. It has been proposed forty years ago that the lepton number can be considered as the fourth ``color" \cite{Pati-Salam-74} in the frameworks based on the gauge group $SU(2)_L \times SU(2)_R \times SU(4)^c$, and the gauge group $SO(10)$ \cite{Fritzsch:1974}, respectively. Right-handed neutrinos are introduced in both models. The descents from the $SO(10)$ are through the $SU(5)$ group or the 
$SU(2)_L \times SU(2)_R \times SU(4)^c$ group. In the latter case the usual color group $SU(3)^c$ is extended to the group $SU(4)^c$ and the leptons are interpreted as the fourth color column. Using the original notations in Ref. \cite{Pati-Salam-74}, the first two generations of fermions are written as   
\begin{equation} \label{4thcolor}
\Psi_{L, R} = \left( \begin{array}{cccc}
\mathcal{P}_a & \mathcal{P}_b & \mathcal{P}_c & \mathcal{P}_d = \nu_e \\ 
\mathcal{N}_a & \mathcal{N}_b & \mathcal{N}_c & \mathcal{N}_d = e^{-} \\ 
\mathcal{\lambda}_a & \mathcal{\lambda}_b & \mathcal{\lambda}_c & \mathcal{\lambda}_d = \mu^{-}\\ 
\mathcal{\chi}_a & \mathcal{\chi}_b & \mathcal{\chi}_c & \mathcal{\chi}_d = \nu_\mu 
\end{array} \right)_{L, R} 
\end{equation}
where $(\mathcal{P}, \mathcal{N}, \mathcal{\lambda}, \mathcal{\chi})$ indicates valency and $(a, b, c, d)$ indicates color degrees of freedom. In Ref. \cite{Pati-Salam-74} the matrices (\ref{4thcolor}) are also written in a composite way, with the more familiar notations
\begin{equation}
\Psi_{L, R} = \left( \begin{array}{cccc}
u_1 & u_2 & u_3 & u_4 = \nu_e \\ 
d_1 & d_2 & d_3 & d_4 = e^{-} \\ 
c_1 & c_2 & c_3 & c_4 = \nu_{\mu} \\ 
s_1 & s_2 & s_3 & s_4 =  \mu^{-} 
\end{array} \right)_{L, R} = \left( \begin{array}{c}
\mathcal{F}_1 \\ 
\mathcal{F}_2 \\ 
\mathcal{F}_3 \\ 
\mathcal{F}_4
\end{array}  \right)_{L, R} \otimes (\mathcal{B}_1, \mathcal{B}_2,  \mathcal{B}_3,  \mathcal{B}_4)
\end{equation}
where $\mathcal{F}$ are spinors carrying the spin of $\Psi$ while $\mathcal{B}$ are scalars carrying the (color) charge of $\Psi$. As mentioned in Ref. \cite{Pati-Salam-74}, it is attractive to consider the components $(\mathcal{F}, \mathcal{B})$ in the decomposition
\begin{equation} \label{FB}
\Psi = \mathcal{F} \mathcal{B}, ~~\mathcal{F} = ( \mathcal{F}_1,  \mathcal{F}_2, \mathcal{F}_3, \mathcal{F}_4 )^T, ~~\mathcal{B} =  ( \mathcal{B}_1,  \mathcal{B}_2, \mathcal{B}_3, \mathcal{B}_4 ),
\end{equation}
as  {\it fundamental fields} and $\Psi$ as {\it composite} ones. Similar ideas has been used in some preon models, for example, the ``halpon" model \cite{Fritzsch:1981} which consider quarks and leptons as bound states of some more fundamental particles (preons) called halpons. In the halpon model the preons are a weak SU(2) doublet of colorless fermions $(\alpha, \beta)$, and a quartet of scalars $(x^i, y), i = R, G, B$ and $y$ carries the fourth color (lepton number), thus leading to an $SU(4)$ symmetry. The first generation of fermions 
\begin{equation} \label{haplon}
\nu = (\alpha y), ~e^{-} = (\beta y), ~ u = (\alpha x), ~d = (\beta x). 
\end{equation}
which has the same spin-charge separation pattern $\Psi = \mathcal{F} \mathcal{B}$  as in Eq. (\ref{FB}). The vector bosons are also composite ones
\begin{eqnarray}
W^{+} &=& (\alpha\bar{\beta}), ~ W^3 = \frac{1}{\sqrt{2}} (\alpha\bar{\alpha} + \beta\bar{\beta}),  \cr
 W^{-} &=& (\bar{\alpha} \beta),  ~Y^0  =  \frac{1}{\sqrt{2}} (\alpha\bar{\alpha} - \beta\bar{\beta}), 
\end{eqnarray}
However,  we will not address here the decomposition of gauge field. The spin-charge separation of non-abelian gauge fields is a more complicated issue (see for example Refs. \cite{Niemi:2005, Faddeev:2006}). For simplicity we may consider them as fundamental fields at least in this paper.

%%%%%%%%%%%%%%%%%%%%%%%%%%%%%%%%%%%%%%%
\section{Spin-charge separation from condensed matter physics}
%%%%%%%%%%%%%%%%%%%%%%%%%%%%%%%%%%%%%%%

In the previous section we have seen how the spin-charge separation may happen in particle physics models.
In condensed matter physics, spin-charge separation describes electrons in some materials as ``bound states" of spinon and chargon (or holons), which carry the spin and charge of electrons respectively. Under certain conditions, such as the high-temperature cuprate superconductivity, the ``composite" electrons can have a deconfinement phase and the spinon and chargon becomes independent particles. Many elaborations of this idea followed in the studies of high $T_c$ superconductors (see e.g. Ref. \cite{Wen:2006} for a review).  
There are experimental observations and computer simulations supporting this idea --- The first direct observations of spinons and holons was reported in Ref. \cite{Shen:2006}; Simulations on spin-charge separation via quantum computing has been performed in Ref. \cite{Kwek:2011}. To demonstrate the basic idea we take the slave-boson formalism in the t-J model as an example \cite{Barnes:1976}.  It is well-known that the low-energy physics of the high-temperature cuprates can be described by the t-J model
\begin{equation}
H = \sum_{ij} J ( {\bf S}_i \cdot {\bf S}_j - \frac{1}{4} n_i n_j ) - \sum_{ij, \sigma} t_{ij} ( c^{\dagger}_{i \sigma} c_{j \sigma} + h.c.),
\end{equation}
where $c^{\dagger}_{i \sigma}, c_{i \sigma}$ are the projected electron operators with constraint $\sum_\sigma c^{\dagger}_{i \sigma} c_{i \sigma} \leqslant 1$, which can be treated with the slave-boson approach by writing 
\begin{equation} \label{cfb}
 c^{\dagger}_{i \sigma} = f^{\dagger}_{i \sigma} b_i 
\end{equation}
where the operators $f^{\dagger}_{i \sigma}$ creates a chargeless spin 1/2 fermion state -- ``spinon"
and $b_i $ creates a charged spin 0 boson state -- ``holon", respectively. Again, Eq. (\ref{cfb}) shows the same decomposition pattern as in Eq. (\ref{FB}) and Eq. (\ref{haplon}) although it is at the operator level. 
 Note that for the (anti)commutator relations to work out correctly a constraint:
\begin{equation} \label{constraint}
f^{\dagger}_{i \uparrow} f_{i \uparrow} +  f^{\dagger}_{i \downarrow} f_{i \downarrow} + b^{\dagger}_i b_i =1
\end{equation}
must be satisfied. What's more, a U(1) gauge symmetry 
\begin{equation}
b_i \rightarrow e^{i \theta} b_i, ~~ f_{i \sigma} \rightarrow e^{i \theta} f_{i \sigma}
\end{equation}
emerges. The d-wave high-$T_c$ superconducting phase appears when holons condense $\left\langle b_i^{\dagger} b_i \right\rangle \neq 0.$  We emphasize on that, as Eq. (\ref{constraint}) suggests, to apply spin-charge separation we need a finite-density system. An isolated electron cannot be decomposed into the spin and charge components. Besides the high $T_c$ cuprate superconductors in the condensed matter physics, the early universe and the inner cores of the compact stars may provided such a finite-density environment in cosmology and astrophysics.

%%%%%%%%%%%%%%%%%%%%%%%%%%%%%%%%%%%%%%%
\section{Spinon and chargon: general couplings}
%%%%%%%%%%%%%%%%%%%%%%%%%%%%%%%%%%%%%%%

The examples from the 4th-color, the halpon model and the high-$T_c$ superconductors tell us that a general spin-charge separation can be written as 
\begin{equation} \label{gFB}
 \Psi = \mathcal{F} \mathcal{B}
\end{equation}
Extra constraint(s) like Eq. (\ref{constraint}) might be needed to form the correct (anti) commutation relations for the operators, and to match the degrees of freedom before and after the spin-charge separation.  For simplicity and for the purpose of studying neutrinos, we will only consider the spin-electric-charge separation. In contrast to most studies of electrons in the condensed matter physics,  in cosmology and particle physics we have a very special type of particles -- the neutrinos. 
Being electrically neutral, neutrinos seem to be the right candidate for the spinon from the spin-charge separation point of view. This will be discussed in the next section.

From the bound state or the confined phase to the deconfined phase in which the spin and charge degrees of freedom are decoupled, it merits some study to see how the effective interaction between spinon and chargon changes. In the confined phase, the component fields $\mathcal{F}$ and $\mathcal{B}$ are coupled, even for a free fermion field $\Psi$. Plugging Eq.(\ref{gFB}) into the kinetic term of a Dirac spinor, 
\begin{equation}
 i \bar{\Psi} \gamma^\mu \overleftrightarrow{\partial}_{\mu} \Psi 
\end{equation}
we obtain an effective coupling between the spinor current and the chargon current
\begin{equation} \label{JJ}
\sim g_{J} ~J_{\tiny{\textrm{spinon}}} \cdot J_{\tiny{\textrm{chargon}}},
\end{equation}
where $g_{J}$ is an effective coupling constant. The spinor current and the chargon current are defined as
\begin{eqnarray}
J^\mu_{\tiny{\textrm{spinon}}} &=&  \bar{\mathcal{F}} \gamma^\mu \mathcal{F} \cr
J^{\mu}_{\tiny{\textrm{chargon}}} &=& \frac{1}{2i}\left(  \mathcal{B}^{\ast}\partial^{\mu}\mathcal{B} -\mathcal{B}\partial^{\mu}\mathcal{B}^{\ast}\right).
\end{eqnarray}
We can do similar spin-charge separation in the Weyl representation
\footnote{Reality condition and chiral projection do not combine in four-dimensional Minkowski spacetime. Therefore one should have two Majorana spinors plus other scalars for the spin-charge separation. Extra degrees of freedoms should be removed by constraints like (\ref{constraint}).}
\begin{equation}
\Psi = |\mathcal{B}_L| e^{i \alpha_L /2} \mathcal{F}_L + |\mathcal{B}_R| e^{i \alpha_R /2} \mathcal{F}_R.
\end{equation}
With constraints like (\ref{constraint}) and $\alpha_L = - \alpha_R$, a Dirac mass term corresponds to a Yukawa-type coupling
\begin{equation} \label{Yukawa}
\bar{\mathcal{F}} (\textrm{Re}\mathcal{B} + i \gamma_5 \textrm{Im}\mathcal{B} ) \mathcal{F}. 
\end{equation}
The Majorana representation and the Majorana mass term will be discussed in details in a separate publication \cite{XC}.

To phenomenologically describe the chargon condensate,  we introduce a non-linear Klein-Gordon equation (for simplicity we only consider the U(1) case)
\begin{equation} \label{Beqt}
\partial^{\mu}\partial_{\mu}  \mathcal{B}
-\lambda\left(  \left\vert \mathcal{B}\right\vert ^{2}-\mathcal{B}_{0}^{2}\right)  \mathcal{B}-i g_J
J_{\tiny{\textrm{spinon}}}^{\mu}\partial_{\mu}\mathcal{B}=0,
\end{equation}
derived from a Higgs potential $V(\mathcal{B}^*, \mathcal{B}) $ and the current-current interaction (\ref{JJ}). 
As the chargon current is proportional to the phase of the $\mathcal{B}$ field ($\mathcal{B} \equiv |\mathcal{B}| e^{i \alpha}$), it is natural to consider the possibility of generating vorticity in the chargon condensate by the spinon current. Here is a simple and interesting example simulating that a rotating spinon current induces vorticity in the chargon condensate \footnote{Eq. (\ref{Beqt}) has been numerically solved in Ref. \cite{Huang:2013, Xiong:2014} and vortex patterns have been observed. Nevertheless, the complex scalar field in Ref. \cite{Huang:2013, Xiong:2014}, being the order parameter for describing a superfluid vacuum, has a different physical meaning from the chargon condensate in the present paper.}  
\begin{equation} \label{local}
J_{\tiny{\textrm{chargon}}}^{\mu}=\left( \rho,\vec{J}\right), ~~\vec{J}=\rho \vec{\Omega} \times \vec{r}
\end{equation}
where $\rho$ is a density distribution and $\Omega$ is an angular velocity. The current-current interaction with (\ref{local}) mimics the local rotation effects of the Coriolis force. It can also simulate the case in which the vorticity is induced by an external magnetic background with an axial symmetry. Fig. 1 shows a three-dimensional vortex-ring lattice as a numerical solution to Eq. (\ref{Beqt}).  
As it can be seen in the next section, such a topological configuration in the chargon condensate can trap spinon zero-mode due to the Yukawa type coupling (\ref{Yukawa}).
 
\begin{figure}[!t]
\begin{center}
\includegraphics[width=3in]{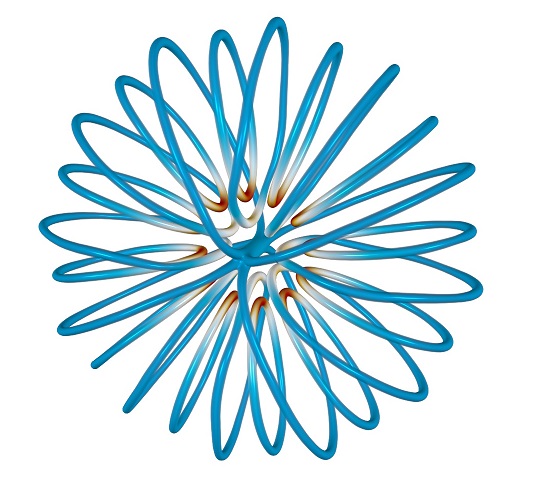}
\caption{A numerical solution to the chargon equation (\ref{Beqt}). It is a three-dimensional vortex ring lattice, induced by the spinon-chargon coupling (\ref{JJ}) (see Refs. \cite{Huang:2013, Xiong:2014} for two-dimensional vortex lattice solutions and more computations on vortex dynamics). Such topological configurations can trap light chiral fermions and release them quickly when the vortex lines split and reconnect. This mechanism may be used to explain the emission of extremely high energy neutrino from cosmic strings.}
\end{center}
\end{figure}

%%%%%%%%%%%%%%%%%%%%%%%%%%%%%%%%%%%%%%%
\section{Neutrinos and spinons}
%%%%%%%%%%%%%%%%%%%%%%%%%%%%%%%%%%%%%%%

Can neutrinos be considered as spinons with respect to the electric-charge-spin separation? It will depend on the material environment. A more meaningful question might be : Is it useful to think neutrinos as spinons from the spin-charge separation?   The answer seems to be yes from the point of view of neutrino condensation \cite{Chodos:1999, Minkowski:2009} and neutrino superfluidity \cite{Kapusta:2004}, as well as their applications in dark matter and other cosmological issues. In Ref. \cite{Kapusta:2004} massive neutrinos are shown to display BCS superfluidity by forming Cooper pairs through the exchange of attractive scalar Higgs boson between left- and right-handed neutrinos. 

Now we study a system where the spinon $\mathcal{F}$ interacts with the chargon $\mathcal{B}$ and another gauge field $A_{\mu}$. \footnote{The gauge field $A_\mu $ is not necessary for the formation of the spinon-string bound state. It is included to the system because in our case the spinon is only neutral to the separated electric charge, it may still carry other quantum numbers like the color charge or weak isospin. When it is minimally coupled to other gauge fields, the current-current coupling (\ref{JJ}) can be absorbed by a gauge transformation, then what is left is the Yukawa type coupling.} The vortex-ring lattice in Fig. 1 is quite complicated for further studies, so we assume that the holon condensate provides a single, straight vortex line background $\mathcal{B} = |\mathcal{B}| (\rho) e^{i m \theta} (m $ is the winding number and $\rho, \theta$ are the polar coordinates in the transverse plane), and $A_{\mu}$ provides a gauge background $A_\mu  = (A_0, A_1, 0, 0)$. 
As the current-current interaction (\ref{JJ}) decouples in the chargon condensation phase and the Yukawa coupling (\ref{Yukawa}) dominates, the system is then reduced to the case studied in Ref.\cite{Callan-Harvey, XC2}.
\footnote{It is reasonable to expect that the current-current interaction (\ref{JJ}) decouples in the deconfined phase, since it comes from the kinetic term of the composite particle which is only meaningful  in the bound state or confined phase. The Yukawa type coupling, however, might survive in the deconfined phase due to the potential nature of the mass term.} The equations of motion of the spinons read (for $m=1$) \cite{Callan-Harvey, XC2}
\begin{eqnarray}  
i \gamma^i (\partial_i - ig A_i) \mathcal{F}_L + i( \gamma^2 \cos \theta  + \gamma^3 \sin \theta ) \partial_{\rho} \mathcal{F}_L  &=& -   |\mathcal{B}|(\rho) e^{- i \theta} \mathcal{F}_R , \cr
i \gamma^i (\partial_i - ig A_i) \mathcal{F}_R + i( \gamma^2 \cos \theta  + \gamma^3 \sin \theta ) \partial_{\rho} \mathcal{F}_R  &=& -  |\mathcal{B}|(\rho) e^{+ i \theta} \mathcal{F}_L, 
\end{eqnarray} 
The spinon zero-mode has an exponential profile \cite{Callan-Harvey, XC2}
\begin{equation} \label{exp}
\mathcal{F}_L = \chi_{L} \, \exp \big[- \int_0^{\rho}  |\mathcal{B}|(\rho') d\rho' \big] 
\end{equation}
and $\mathcal{F}_R = i \gamma^2 \mathcal{F}_L$.
$\chi_{L}$ is a two-dimensional spinor satisfying 
\begin{equation}
i \gamma^i (\partial_i - ig A_i) \chi_{L} = 0
\end{equation} 
The exponential profile of $\mathcal{F}_L$ in Eq. (\ref{exp}) shows that chiral zero-modes of spinons are localized on a vortex, which can be considered as a fermion-vortex bound-state. If we take the spinons to be the neutrinos, and think about more complicated configurations than the one shown in Fig.1, such as a vortex tangle or quantum turbulence in the chargon condensate, then the dynamical vortex lines or rings will split and reconnect, the trapped neutrinos will be emitted quickly during the process of reconnection. The resulting phenomena might be similar to the cases in which cosmic strings emit extremely high energy neutrinos \cite{Lunardini:2012}.   
Note that such mechanism has been used in the scalar field cosmology \cite{Huang:2012, Huang:2013, Xiong:2014} where a superfluid cosmological vacuum plays the role of the chargon condensate here.

\section{Conclusions and discussions}

Inspired by the spin-charge separation scenario in difference physics models, such as the fourth-color model, the haplon model, and the t-J model in particle physics or condensed matter physics, we give a general picture about how the composite models split into spin and charge sectors under certain environment condition (e.g. finite density), and derive some simple but probably universal couplings. This provides a new understanding about the lepton-quark unification and suggests the similarity between the spinons and the neutrinos (with respect to the electric-charge-spin separation). We also give a simple phenomenological description for the chargon condensate, and show that it supports vorticity, which leads to the localization of the spinons and hence the formation of spinon-string bound states. This may provide an alternative explanation for the emission of extremely high energy neutrinos from cosmic strings.

\section*{Acknowledgement}

I thank Peter Minkowski, Harald Fritzsch and Kerson Huang for valuable discussions, and Xiaopei Liu and Yulong Guo for their support on the simulation and the visualization of superfluid vorticity. This work is supported by the research funds from the Institute of Advanced Studies, Nanyang Technological University, Singapore.

\bibliographystyle{ws-procs975x65}
\bibliography{ws-pro-sample}

%Non BiBTeX users can list down their references as:

\end{document}